\documentclass[11pt]{article}
\usepackage{my_moriond,psfig}

\def\Journal#1#2#3#4{{#1} {\bf #2}, #3 (#4)}

\def\be{\begin{equation}}
\def\ee{\end{equation}}
\def\bea{\begin{eqnarray}}
\def\eea{\end{eqnarray}}

\def\la{\mathrel{\vcenter{\offinterlineskip\halign{\hfil$##$
\hfil\cr<\cr\sim\cr}}} \!\!}

\newcommand{\qniec}{\end{document}}
\newcommand{\etal}{{\it et al.}}
\newcommand{\de}{\delta}
\newcommand{\te}{\theta}

\newcommand{\f}{\frac}
\newcommand{\s}{\sigma}
\newcommand{\bfx}{\mathbf{x}}

\newcommand{\bfv}{\mathbf{v}}

\newcommand{\calO}{{\mathcal O}}

\newcommand{\calK}{{\mathcal K}}

\newcommand{\bs}{\begin{slide}}
\newcommand{\bc}{\begin{center}}
\newcommand{\ec}{\end{center}}
\newcommand{\es}{\end{slide}}
\newcommand{\lan}{\langle}
\newcommand{\ran}{\rangle}

\newcommand{\hmpc}{$h^{-1}\,\mathrm{Mpc}$}

\def\lta{\mathrel{\spose{\lower 3pt\hbox{$\mathchar"218$}}
 \raise 2.0pt\hbox{$\mathchar"13C$}}}

\begin{document}
%\vspace*{4cm}
\title{LARGE-SCALE DENSITY--VELOCITY RELATIONS}

\author{ M.J. CHODOROWSKI }

\address{Copernicus Astronomical Center, Bartycka 18, 
	00--716 Warsaw, Poland}

\maketitle\abstracts{I present recent progress in theoretical 
modelling of cosmological density--velocity relations in the weakly
nonlinear regime. The relations are local, based on rigorous
perturbation theory and include the effects of smoothing of the
density and the velocity fields. For small smoothing scales, they can
be improved by slight adjustments based on N-body results. The
relations can be useful for density--velocity comparisons, yielding a
fair estimate of $\Omega$ and offering a method for disentangling
$\Omega$ and bias.  }

\section{Introduction} \label{sec-intro}

In the gravitational instability paradigm for the formation of
structure in the Universe, the peculiar motions of galaxies are
tightly related to the large-scale mass distribution. The comparison
between the density and the velocity fields can serve as a test of the
gravitational instability hypothesis and as a method for estimating
the cosmological parameter $\Omega$ (Dekel \etal~\cite{dek93}). In
linear theory, the relation between the density and the velocity
fields is
\be 
\de(\bfx) = - f(\Omega)^{-1} \mathbf{\nabla}
\cdot \bfv(\bfx) \,,
\label{e1}
\ee 
where $f(\Omega) \simeq \Omega^{0.6}$. The relation between the
density contrast and the velocity divergence is thus linear and local.

This equation is applicable only when the density fluctuations are
small compared to unity, $\s_\de^2 \equiv \lan \de^2 \ran \ll
1$. However, sampling of galaxies in current redshift surveys and
random errors in the peculiar velocity data enable reliable dynamical
analysis at smoothing scales down to a few \hmpc, where fluctuations
already exceed the regime of applicability of linear theory. On the
other hand, even at scales as small as a few \hmpc, `typical' (rms)
fluctuations are still not in excess of unity.\footnote{For a Gaussian
smoothing function, commonly used in the density--velocity
comparisons.} One could thus hope that to construct the relation
between the fields in question, perturbation theory can be effectively
used.

In perturbation theory, one approximates the evolved density contrast
as a sum of terms $\de_j$, each corresponding to the $j^{th}$ order
in perturbation theory,
\be 
\de = \de_{1} + \de_{2} + \de_{3} + \ldots \,.
\label{e4}
\ee
Similarly, defining a variable $\te$ proportional to the velocity
divergence,
\be
\te = - f(\Omega)^{-1} \mathbf{\nabla}\cdot\bfv \,,
\label{e3}
\ee 
we can expand it in a perturbative series,
\be 
\te = \te_{1} + \te_{2} + \te_{3} + \ldots \,.
\label{e5}
\ee
The solutions for $\de_j$ and $\te_j$ were obtained more than a decade
ago (Fry~\cite{fry}, Goroff \etal~\cite{gor}). In general, the
$j^{th}$ order solution is of the order of $(\de_1)^j \sim
\s_\de^j$. In the weakly nonlinear regime, i.e., when $\s_\de < 1$, 
which is our case, we may hope the perturbative series to converge
rapidly.

Linear theory is nothing but perturbation theory truncated at the
lowest, i.e.\ the first order. A natural way of extending the linear
relation~(\ref{e1}) is to include higher-order contributions to the
density contrast and the velocity divergence. Unfortunately, these
contributions are non-local, hence the nonlinear density vs
velocity-divergence relation (DVDR) \emph{at a given point} has a
scatter. Therefore, as a local estimator of the density from the
velocity divergence we adopt the conditional mean, i.e., the mean
density \emph{given} the velocity divergence.

The calculation of the conditional moments is a statistical
problem. To compute mean $\de$ given $\te$, $\lan \de \ran|_{\te}$, we
need the joint probability distribution function (PDF) $p(\de,\te)$.
Under an assumption of Gaussian initial conditions, adopted in this
work, this function is initially a bivariate Gaussian (degenerated,
since $\de_1 = \te_1$). Subsequent nonlinear gravitational evolution,
however, drives the PDF away from its initial shape. Our problem
here is to calculate it in the weakly nonlinear regime.

Mathematically, a PDF is given by the inverse Fourier transform of
its characteristic function $\Phi$, which in turn is related to the
cumulant generating function $\calK$ by the equation $\Phi =
\exp{[\calK]}$. The cumulants, $\kappa_{mn}$, from which $\calK$ is
constructed,

\be
\calK(it,is) = \sum_{(m,n) \ne (0,0)}^\infty  
\f{\kappa_{mn}}{m! n!}
(it)^m (is)^n \,, 
\label{e7}
\ee
are the \emph{connected} (reduced) part of the joint moments of $\de$
and $\te$, $\kappa_{mn} = \lan \de^m \te^n \ran_{\rm conn}$. We
introduce $\calK$ because in the weakly nonlinear regime, the
cumulants obey a scaling hierarchy in $\s_\de$ such that the
series~(\ref{e7}) is a power series in $\s_\de$ (even in the case of
the standard variables $\mu \equiv \de/\s_\de$ and $\nu \equiv
\te/\s_\te$). Therefore, truncating~(\ref{e7}) at order $p$ we neglect
contributions which are $\calO(\s_\de^{p+1})$. This yields the
truncated PDF and in turn the weakly nonlinear DVDR up to a given
order in perturbation theory.

\section{Results} \label{sec-results}
 
Up to third order (Chodorowski \& {\L}okas~\cite{cl97}) we have
\be
\lan \de \ran|_{\te} = a_1 \te + a_2 (\te^2 - \s_\te^2) + a_3 \te^3 \,,
\label{e10}
\ee
where $\s_\te^2$ is the variance of the velocity divergence field. 
The reverse relation (Chodorowski \etal~\cite{chod98}) is
\be
\lan \te \ran|_{\de} = r_1 \de + r_2 (\de^2 - \s_\de^2) + r_3 \de^3 
\label{e11}
\ee
(an extension of a second-order formula by
Bernardeau~\cite{b92}). Because the DVDR has a scatter,
expression~(\ref{e11}) cannot in general be obtained by direct
inversion of expression~(\ref{e10}). In density--density comparisons,
one reconstructs the mass density field from the observed velocity
field and compares it to the observed galaxy density field. In
velocity--velocity comparisons, one reconstructs (under some
assumptions about bias) the velocity field from the observed galaxy
density field and compares to the observed velocity
field. Formula~(\ref{e10}) is thus relevant for density--density
comparisons, while formula~(\ref{e11}) is relevant for
velocity--velocity comparisons. Indeed, for irrotational flows the
velocity field can readily be recovered from its divergence given some
boundary conditions at large distances,

\be
\bfv(\bfx) = \f{f(\Omega)}{4\pi} \int {\rm d}^3 x' \te(\bfx')
\f{\bfx'- \bfx}{\vert \bfx'- \bfx \vert^3} \,.
\label{e11a}
\ee 

The coefficients $a_j$ and $r_j$ are the previria\-li\-za\-tion-,
skewness-, and kurtosis-like combinations of the joint moments of
$\de$ and $\te$. Calculations of the reduced moments have already
become a small industry in perturbation theory. For Gaussian
smoothing, the related moments for $\de$ and for $\te$ (separately)
were computed by {\L}okas \etal~\cite{lok95} and {\L}okas
\etal~\cite{lok96} The coefficients $a_j$ and $r_j$, though
constructed from the joint moments, possess similar mathematical
structure and were computed in an analogous way.~\cite{cl97,chod98} 

Chodorowski \etal~\cite{chod98} also calculated the scatter in the
DVDR and found it to be relatively small.\footnote{The ratio of the
scatter to the rms fluctuation of the density field, $\s_\de$,
vanishes in the limit $\s_\de \to 0$, as expected.} Nevertheless, one
can try to further reduce the variance of density estimators by
including off-diagonal components of the velocity deformation
tensor. Indeed, including the shear of the velocity field reduces the
scatter in the density--velocity relation.~\cite{man95,chod97}

\section{Tests against N-body simulations} \label{sec-nbody}

\begin{figure}
  \begin{center}
  	\begin{tabular}{lcr}
		\hspace{-23pt} \psfig{figure=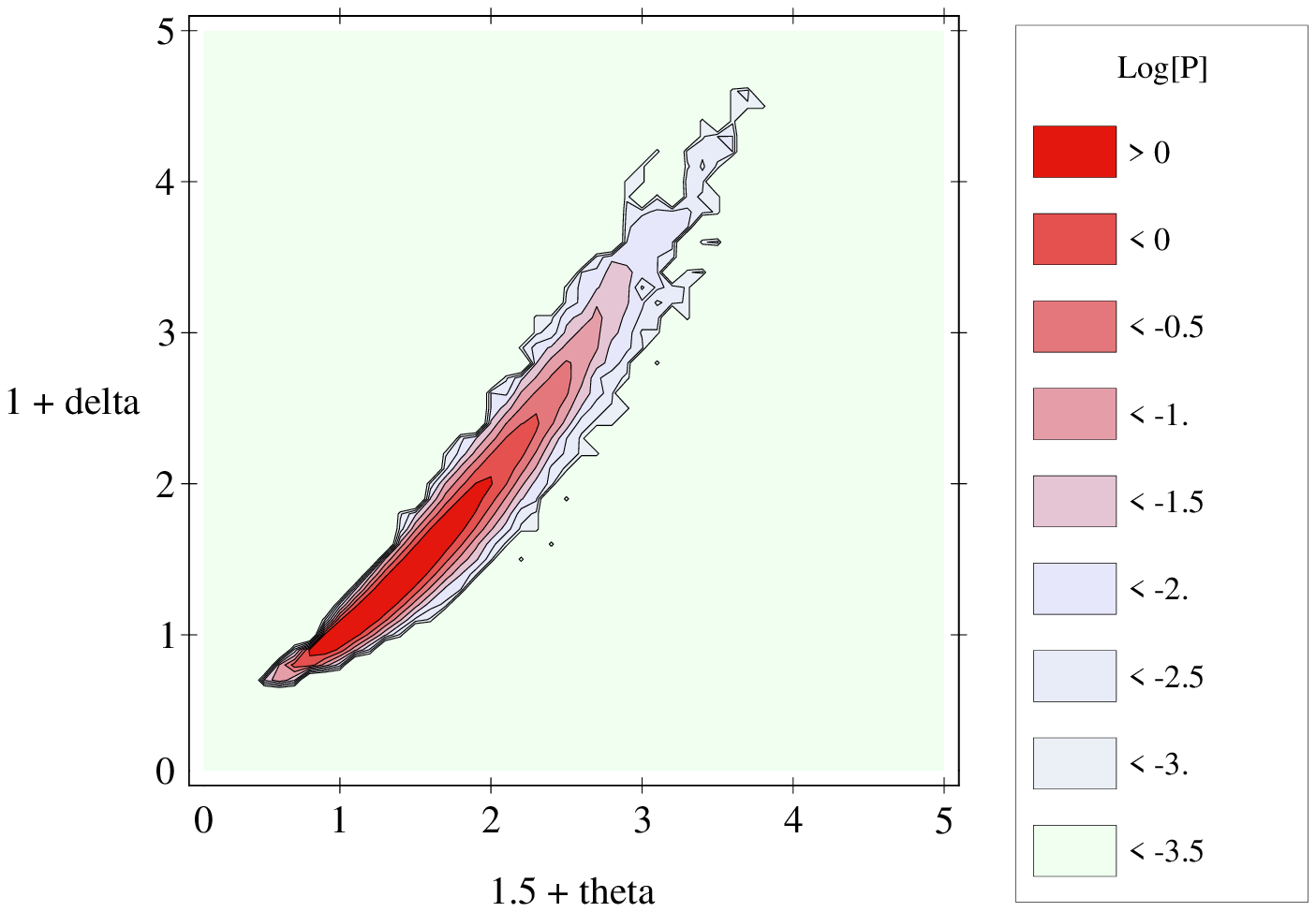,height=2.424in} &
		\hspace{0pt}      		      &
		\psfig{figure=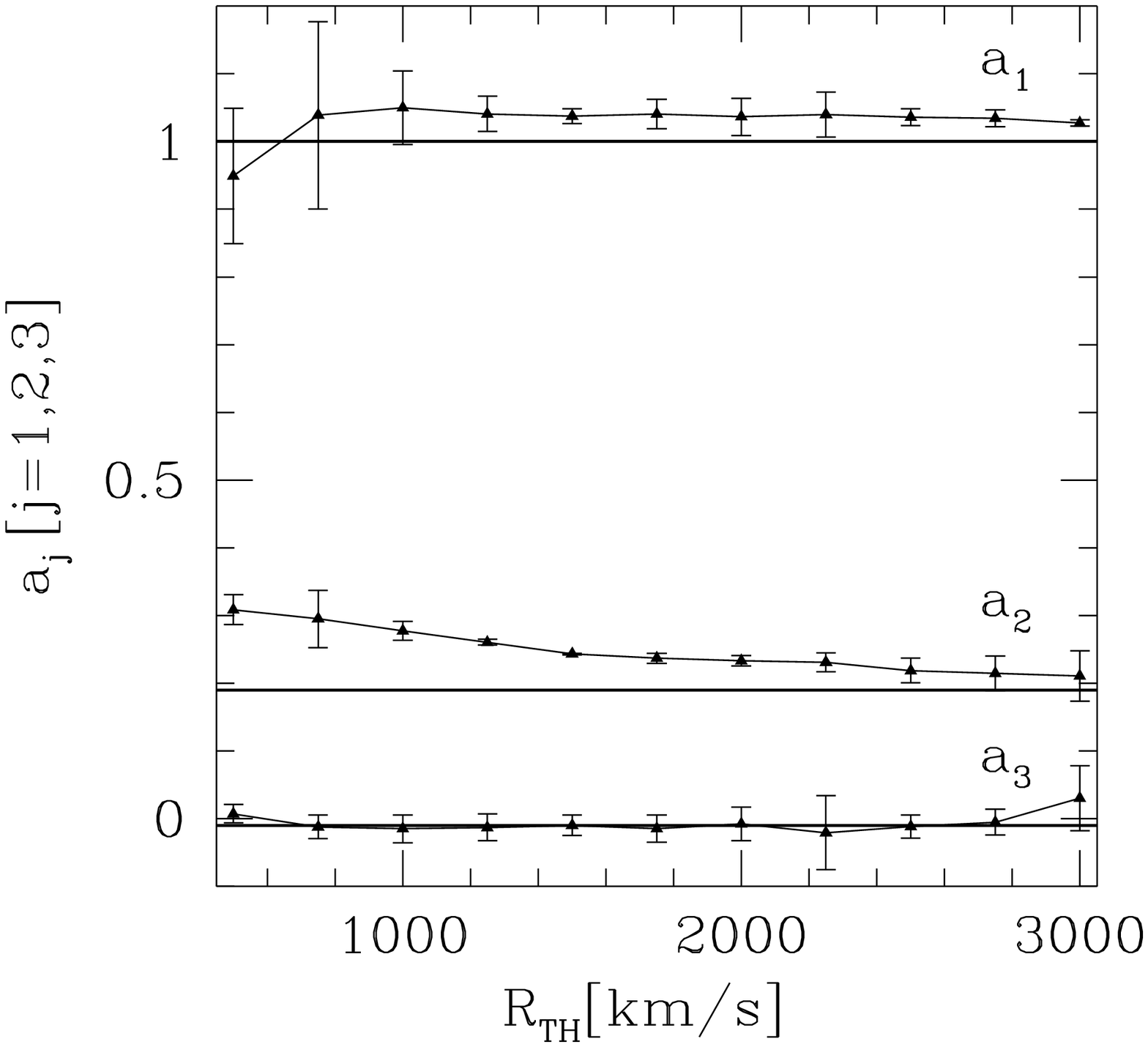,height=2.4in}
  	\end{tabular}
  \end{center}

  \caption{Left: joint probability distribution function for the
    density contrast ($\de$) and the velocity divergence ($\te$) from
    an N-body simulation by Chodorowski \etal,~\protect\cite{cblk98}
    shown against $1 + \de$ and $1.5 + \te$. Right: the coefficients
    $a_j$ (see eq.~[\protect\ref{e10}]), as functions of the smoothing
    scale, from N-body simulations by Chodorowski \&
    Stompor.~\protect\cite{cs98} Simulation results are shown as thin
    curves; third order perturbation theory predictions are shown as
    thick lines. The simulations are for power-law power spectra with
    the spectral index $n$ equal to $-1.5$ (left) and $-1$ (right). A
    top-hat smoothing function is used.}
\label{fig-artwork}
\end{figure}

Perturbative expressions for density--velocity relations should be
checked against N-body simulations, in order to assess their range of
applicability. The results of the simulations by Chodorowski
\etal~\cite{cblk98} and Chodorowski \& Stompor~\cite{cs98} 
(see Figure~\ref{fig-artwork}) can be summarized as follows:

\begin{itemize}

\item \emph{For large smoothing scales} ($\s_\de \ll 1$), the 
numerical coefficients $a_j$ and $r_j$ converge to the predicted
values. (In this sense, the predicted DVDRs are `asymptotically
unbiased'.)

\item \emph{For smoothing scales as small as a few megaparsecs}
($\s_\de \la 1$),
	\begin{itemize} 

	\item The numerical coefficients slightly, but systematically, 
	deviate from predicted values. (Apparently, higher--than
	third--order contributions become significant.)

	\item On the other hand, third-order polynomials, with fitted
	coefficients, provide excellent fits to N-body data.
	(Higher-order contributions modify the values of the
	coefficients for $j = 1,2,3$, but do not significantly induce
	non-zero values for $j >3$.)

	\end{itemize}
\end{itemize}

\section{Implications for the value of $\Omega$} \label{sec-omega}

For the sake of simplicity, here we will consider only
second-order corrections to the linear relation~(\ref{e1}). Neglecting
third-order terms in equation~(\ref{e11}) yields
\be
\te = \de - a_2 (\de^2 - \s_\de^2)
\label{e12}
\ee
($r_2 = - a_2$, $0.2 < a_2 < 0.3$~\cite{chod98,cl97}). Assuming
linear bias between the galaxy density field, $\de_g$, and the mass
density field, $\de_g = b \de$, we can express the relation in terms
of the observable quantities $\mathbf{\nabla}\cdot\bfv$ and $\de_g$,

\be
- \mathbf{\nabla}\cdot\bfv = \beta \de_g - a_2 b^{-1} \beta \de_g^2
\,.
\label{e13}
\ee
Here, $\beta(\Omega,b) \equiv f(\Omega) / b$ and we also neglect the
offset in the relation. We see that the quadratic correction has the
opposite sign to the linear term. It implies that the linear relation
overestimates the predicted velocities. Hence, in order to match the
observed ones, a lower value of $\beta$ is needed. In effect,
\emph{applying linear theory biases the estimate of $\Omega$ low}
(cf.\ Strauss \& Willick~\cite{sw95}). 

Note that the two coefficients of the binomial in $\de_g$ on the RHS
of equation~(\ref{e13}) are different combinations of $\beta$ and $b$,
or $f(\Omega)$ and $b$. This offers a way to solve for $\Omega$ and
$b$ separately.~\cite{cl97}

\section{Summary}
The weakly nonlinear estimators of density from velocity, and vice
versa, presented here,
\begin{itemize}
	\item are asymptotically unbiased;
	\item for smaller smoothing scales, can be improved by slight
	adjustments from N-body data;
	\item should be used in density--density and
	velocity--velocity comparisons, in order not to bias the
	estimate of $\Omega$ low;
	\item offer a method for disentangling $\Omega$ and bias.
\end{itemize}

\section*{Acknowledgments}
This research has been supported in part by the Polish State Committee
for Scientific Research grants No.~2.P03D.008.13 and 2.P03D.004.13.

\end{document}